\documentstyle[aps,preprint]{revtex}
\begin{document}
\tighten
\newcommand{\mb}[1]{\mbox{\boldmath $#1$}}
 
\title{\bf \large
Plateaux Transitions in the Pairing Model:
Topology and Selection Rule
}
\author{
Y. Morita$^1$ and
Y. Hatsugai$^{1,2}$}
\address{
Department of Applied Physics,
University of Tokyo,
7-3-1 Hongo Bunkyo-ku, Tokyo 113, Japan$^1$ \\
PRESTO, Japan Science and Technology Corporation$^{2}$
}
\date{July 1, 1999}
\maketitle

\begin{abstract}
Based on the two-dimensional
lattice fermion model,
we discuss
transitions between different pairing states.
Each phase is labeled by 
an integer
which is a topological invariant
and characterized by
{\it vortices}
of the Bloch wavefunction.
The transitions between phases with different integers
obey a selection rule.
Basic properties of the edge states are revealed.
They reflect the topological character of the bulk.
Transitions driven by randomness are 
also discussed numerically.
\end{abstract}
\narrowtext
Quantum phase transitions
between different superconducting states
have attracted much interest recently.
In refs.\cite{ong1,laughlin1},
for example,
its possible realization
in a high-$T_{c}$ superconductor
was proposed,
which is
accompanied by
the time-reversal symmetry $\cal T$ breaking.
Further,
there is a recent observation
that it has some similarity to
the plateaux transition 
in the integer quantum Hall effect (IQHE)
\cite{volovik1,kagalovsky1,senthil1,gruzberg1}.
One of the claims is that each phase
is labeled by an integer
(an analogue of the Hall conductance 
in the IQHE) and 
there can be
transitions
between phases with different integers.

In this paper,
based on the lattice fermion model,
we investigate the problem.
The integer for each phase
is defined by a topological invariant
of the U(1) fiber bundle 
(the Chern number)
\cite{volovik1,tknn1,yh1}.
The U(1) fiber bundle 
is a geometrical object
which is 
composed of 
the Brillouin zone (torus)
and the Bloch wavefunctions (fiber).
Due to its topological stability,
a singularity in the U(1) fiber bundle
necessarily occurs
with the change of the Chern number.
The singularity is identified with
the energy-gap closing
\cite{hk0,hk1,oshikawa1,hk2,hk3,hk4}.
The Chern number
is closely related to 
the zero points ({\it vortices})
of the Bloch wavefunction.
Focusing on the motion of the vortices
near the singularity,
we give a general proof of
a {\it selection rule} of the transitions.
Due to the intrinsic symmetry of the system,
the selection rule differs
from that of the IQHE\cite{klz1}.
We also investigate
the properties of the edge states 
and how they reflect
the topological character of the bulk.
The transition due to the change of randomness strength
is a typical example of the problem.
As emphasized in ref. \cite{az1},
the symmetry effect
leads to 
a new universality class 
and it is interesting 
as the Anderson localization problem.
We also discuss
the disorder-driven transition
numerically.

The Hamiltonian is
\begin{eqnarray}
{\cal H}=
\sum_{l,m}
{\bf c}_{l}^{\dagger}{\bf H}_{lm}{\bf c}_{m}
=
\sum_{l,m}
{\bf c}_{l}^{\dagger}
\pmatrix{
t_{lm}&\Delta_{lm}\cr
\Delta_{ml}^{*}&-t_{ml}\cr}{\bf c}_{m}
\label{hm1}
\end{eqnarray}
where
${\bf c}_{n}^{\dagger}=
\pmatrix{c_{n\uparrow}^{\dagger},c_{n\downarrow}^{\dagger}\cr}$,
${\bf c}_{n}=
{}^{t}
\pmatrix{c_{n\uparrow},c_{n\downarrow}\cr}$
and $n=(n_{x},n_{y}){\in}{\bf Z}^{2}$.
This is an extension of the lattice fermion model
discussed in connection with
the plateaux transition in the IQHE
\cite{hk1,oshikawa1,hk2,hk3,hk4}.
Here we comment on
the relation between
this Hamiltonian and the superconductivity.
Under the 
unitary transformation
$c_{n\uparrow}{\rightarrow}d_{n\uparrow}$,
$c_{n\downarrow}{\rightarrow}d_{n\downarrow}^{\dagger}$
(for ${\forall}n$),
the Hamiltonian (\ref{hm1})
is equivalent to
${\cal H}=
\sum_{l,m}
[d_{l\uparrow}^{\dagger}{t_{lm}}d_{m\uparrow}+
d_{l\downarrow}^{\dagger}{t_{lm}}d_{m\downarrow}+
d_{l\uparrow}^{\dagger}{\Delta_{lm}}d_{m\downarrow}^{\dagger}+
d_{m\downarrow}{\Delta_{lm}^{*}}d_{l\uparrow}].$
This is the pairing model
for the singlet superconductivity.
In the context of 
the superconductivity,
the pair potential ${\Delta_{lm}}$ should be determined by
the self-consistent equation.
Although the effect is interesting in itself,
it is beyond the scope of this paper.
Further,
the conditions
$t_{lm}^{*}=t_{ml}$ 
and $\Delta_{lm}=\Delta_{ml}$
are imposed and
they correspond to
the hermiticity and the SU(2) symmetry respectively.
The SU(2) symmetry
leads to the condition
\cite{senthil1}
\begin{eqnarray}
-(\sigma_{y}{\bf H}_{lm}\sigma_{y})^{*}
={\bf H}_{lm}.
\label{su2-rl}
\end{eqnarray}
Due to the SU(2) symmetry,
we can restrict ourselves 
to the sector 
$\sum_{n}
{\bf d}_{n}^{\dagger}{\sigma}^{z}
{\bf d}_{n}=0$
without loss of generality.
This is equivalent to 
the {\it half-filled condition}
for the Hamiltonian (\ref{hm1}),
which we impose in the following arguments.

Now let us define
a topological invariant (the Chern number)
for our model.
It is a key concept in the following arguments.
Put the system on a torus,
which is $L_{x}{\times}L_{y}$ and periodic
in both $x$ and $y$ directions.
Define the Fourier transformation by
${\bf c}_{n}=
{1/\sqrt{L_{x}L_{y}}}
{\sum_{k}}
e^{ikn}
{\bf c}(k)$
where
$k=(k_{x},k_{y})$ 
is on the Brillouin zone
$(-\pi,\pi]{\times}(-\pi,\pi]$.
Assuming 
${\bf H}_{lm}$ to be invariant
under translations,
we obtain
${\cal H}
=
{\sum_{k}}
{\bf c}(k)^{\dagger}{\bf H}(k)
{\bf c}(k)$
where
${\bf H}(k)=
{\sum_{(l-m)}}
e^{-ik(l-m)}{\bf H}_{lm}$.
The ${\bf H}(k)$
has two eigenvectors and eigenvalues.
They correspond to 
the Bloch wavefunctions and 
the energy bands respectively.
To satisfy the half-filled condition,
the lower band is occupied for the ground state.
We denote 
the Bloch wavefunction for the lower band
by ${}^{t}\pmatrix{a(k),b(k)\cr}$.
Then
the topological invariant
(the Chern number of the U(1) fibre bundle)
is defined as:
\begin{eqnarray} 
C
=\frac 1 {2\pi i}
\int dk\; \hat z \cdot (\mb{ \nabla}_k \times{\mb A})
\label{cn1} 
\end{eqnarray}
where
${\mb A}= 
\pmatrix
{a^{*}(k),b^{*}(k)\cr}
\mb{\nabla} _k 
{}^{t}
\pmatrix
{a(k),b(k)\cr}$
and
$\hat z=(0,0,1)$
\cite{volovik1,tknn1,yh1}.
The integration $\int dk$
is over the Brillouin zone
which can be identified with a torus.
For simplicity,
we assumed ${\bf H}_{lm}$
to be invariant under translations.
However
a generalization to
a multi-band system 
(including a random system),
is possible.
It is crucial for the following arguments
to rewrite the above formula
in terms of 
a zero point of the Bloch wavefunction ({\it vortex})
and the winding number ({\it charge}).
To be explicit,
let us perform the {\it gauge fixing}
of the Bloch wavefunction 
for the lower energy band.
We note that 
the Chern number itself
does not depend on the gauge fixing.
To
define
the gauge,
we use
the rule $a(k)=1$
and introduce a notation
$b(k)=b'(k)e^{-i\zeta(k)}$
($b'(k){\in}{\bf R}$).
An ambiguity
in the gauge fixing occurs 
when $a(k)=0$.
Around the zero point (vortex)
in the Brillouin zone,
it is necessary
to change the way of the gauge fixing,
for example, as $b(k)=1$.
Then 
the Chern number (\ref{cn1}) is rewritten as
\begin{eqnarray}
C=
\sum_{l} C_{l},
{\ }{\ }
C_{l}=
1/{2\pi} 
\oint_{\partial R_{l}}dk 
{\bf{\nabla}}\zeta(k)
\label{cn2}
\end{eqnarray}
where 
the summation is taken over
all vortices of $a(k)$
and
$R_{l}$ is a region
surrounding the $l$-th vortex
which does not contain
other zeros
of either $a(k)$ or $b(k)$.
Here $C_{l}$ is an integer and 
we call it the charge 
of the $l$-th vortex.

Let us discuss
the $d_{x^{2}-y^{2}}+id_{xy}$ model
on a torus
as an example
\cite{volovik1,senthil1}.
The model is defined by
$t_{n+e_{x},n}
=t_{n+e_{y},n}
=t$,
${\Delta}_{n+e_{x},n}
=-{\Delta}_{n+e_{y},n}
={\Delta}_{x^{2}-y^{2}}$,
${\Delta}_{n+e_{x}+e_{y},n}
=-{\Delta}_{n-e_{x}+e_{y},n}
=i{\Delta}_{xy}$ 
for $\forall n$
($e_{x}=(1,0),e_{y}=(0,1),
t>0,{\Delta}_{x^{2}-y^{2}},{\Delta}_{xy}{\in}{\bf R}$)
and the other matrix elements are zero.
Then 
${\bf H}(k)=
\pmatrix{
A(k)&B(k)\cr
B^{*}(k)&-A(k)\cr}$
where
$A(k)=2t(\cos{k_{x}}+\cos{k_{y}})$
and 
$B(k)=
2{{\Delta}_{x^{2}-y^{2}}}
(\cos{k_{x}}-\cos{k_{y}})
+
2i
{{\Delta}_{xy}}
[\cos{(k_{x}+k_{y})}-\cos{(k_{x}-k_{y})}]$.
The energy spectrum is given by
$E={\pm}{\sqrt{A(k)^{2}+|B(k)|^{2}}}$.
When ${\Delta}_{xy}=0$,
the upper band and the lower band
touch at four points 
$(\pm{\pi/2},\pm{\pi/2})$
in the Brillouin zone.
The low-lying excitations 
around the gap-closing points
are described by
massless Dirac fermions.
By turning on a finite ${\Delta}_{xy}$,
the mass generation occurs in the Dirac fermions.
The vortex position 
is given by $a(k)=0$ and
it is $(k_{x},k_{y})=(0,0)$.
Using
$B(k)=B'(k)e^{i\zeta(k)}{\ }
(B'(k){\in}{\bf R})$,
the charge of the vortex is 
$1/{2\pi}
\oint_{(0,0)}dk 
{\bf{\nabla}}{\zeta (k)}=2{\ }
{\rm sgn}
({\Delta_{xy}}/{\Delta_{x^{2}-y^{2}}})$.
Since there is no other vortex,
the Chern number $C$ is given by
\begin{eqnarray}
C=
2{\ }{\rm sgn}
({\Delta_{xy}}/{\Delta_{x^{2}-y^{2}}}).
\label{chd+d}
\end{eqnarray}
Here we note that,
as in the case of the IQHE on the lattice, 
the Chern number can
take various integer values in general cases
e.g. a multi-band system and 
a model with a different $\cal T$-broken pairing symmetry.

As in the QHE
\cite{halperin1,hatsugai1,wen1},
the {\it edge states} 
play a crucial role in the problem
\cite{volovik1,kagalovsky1,senthil1}.
The edge states reflect the bulk properties
and
it is possible to detect
the topological character of the bulk
through the edge states.
In order to discuss the edge states,
put the system on a cylinder
which is $L_{x}{\times}L_{y}$ and periodic
only in $y$ direction.
Further
we impose an open boundary condition
in $x$ direction.
Define the Fourier transformation by
${\bf c}_{n}=
{1/\sqrt{L_{y}}}
{\sum_{k_{y}}}
e^{ik_{y}n_{y}}
{\bf c}_{n_{x}}(k_{y})$
where
$k_{y}$ is on $(-\pi,\pi]$.
Then
${\cal H}
=
{\sum_{k_{y}}}
{\bf c}_{l_{x}}(k_{y})^{\dagger}
{\bf H}_{l_{x}m_{x}}(k_{y})
{\bf c}_{m_{x}}(k_{y})$
where
${\bf H}_{lm}$
is assumed to be invariant
under translations
in $y$ direction and
${\bf H}_{l_{x}m_{x}}(k_{y})=
{\sum_{(l_{y}-m_{y})}}
e^{-ik_{y}(l_{y}-m_{y})}{\bf H}_{lm}$.
The relation (\ref{su2-rl}) 
can be rewritten as
$-(\sigma_{y}{\bf H}_{l_{x}m_{x}}(k_{y})\sigma_{y})^{*}
={\bf H}_{l_{x}m_{x}}(-k_{y})$.
Define 
an eigenvector ${\bf u}$
by
$\sum_{m_{x}}
{\bf H}_{l_{x}m_{x}}(k_{y})
{\bf u}_{m_{x}}
=E
{\bf u}_{l_{x}}$.
Now we show that
there are two basic operations
$\cal P$ and $\cal Q$
on the vector.
They are defined by
$
({\cal P}{\bf u})_{n_{x}}
=(\sigma_{y}{\bf u}_{n_{x}})^{*}$
and
$
({\cal Q}{\bf u})_{n_{x}}
={\bf u}_{L_{x}-n_{x}+1}$.
Then
$\sum_{m_{x}}
{\bf H}_{l_{x}m_{x}}(-k_{y})
({\cal P}{\bf u})_{m_{x}}=
(-E)
({\cal P}{\bf u})_{l_{x}}$.
Further,
when $t_{lm}$ is real and uniform,
we can obtain another relation
$\sum_{m_{x}}
{\bf H}_{l_{x}m_{x}}(-k_{y})
({\cal Q}{\bf u})_{m_{x}}=E
({\cal Q}{\bf u})_{l_{x}}$.
Based on the symmetry, 
we shall discuss basic properties 
of the edge states.
Consider a case when
${\bf u}$ is an eigenstate 
which is localized spatially
on the left (or right) boundary
i.e. a left (or right)-hand edge state.
Then, from the above argument,
${\bf u}$,${\cal P}{\bf u}$,${\cal Q}{\bf u}$,${\cal PQ}{\bf u}$
are classified into 
two left-hand edge states and two right-hand edge states.
Now, 
as in the argument of the IQHE\cite{laughlin2},
let us introduce a fictitious flux through the cylinder
and change it from $0$ to flux quanta $hc/e$.
Due to the symmetry,
the number of edge states 
which move
from one boundary to the other is necessarily {\it even}.
We shall consider the
$d_{x^{2}-y^{2}}+id_{xy}$ model
on a cylinder
as an example.
In Fig.1, 
the energy spectrum is shown.
It can be confirmed that
the energy spectra of the edge states 
appear in pairs and
the number of the edge states
which move from one boundary to the other
as the fictitious flux is added,
is even and
coincides with the Chern number.
In other words,
the edge states 
directly reflect
the topological character of the bulk.

As discussed above, 
each phase in our model is labeled 
by the Chern number.
The vortices move
as the Hamiltonian is perturbed.
The Chern number, however, does not change in general.
Due to the topological stability,
the change of the Chern number 
is necessarily 
accompanied by
a singularity in the U(1) fiber bundle.
The singularity is identified with 
the energy-gap closing.
Focusing on the singularity, we can prove
a {\it selection rule} 
from a general point of view
(see also \cite{hk1,oshikawa1,hk2,hk3,hk4}).
The selection rule is closely
related to the SU(2) symmetry in our model.
Let us introduce a parameter $g$
in the Hamiltonian.
Assume that,
when $g=g_{0}$,
the energy gap closes
at several zero-energy points
in the Brillouin zone.
Next focus on 
the region near
one of the gap-closing points 
$(k_{x}^{0},k_{y}^{0})$ i.e. 
${\bf{p}}=
{}^{t}\pmatrix{k_{x}-k_{x}^{0}, k_{y}-k_{y}^{0}, g-g_{0}}
\sim{\bf 0}$.
Then the leading part of the Hamiltonian is, generally,
given by
\begin{eqnarray}
{\bf H}_{0}({\bf p})
=
1{\bf v}_{0}{\bf p}
+
({\sigma}_{x},{\sigma}_{y},{\sigma}_{z})
\bf{v}
\bf{p}
\label{osieq1}
\end{eqnarray}
where 
${\bf v}_{0}$ is a $1{\times}3$ vector,
${\sigma}_{x(y,z)}$ is a  $2{\times}2$ Pauli matrix 
and ${\bf v}$ is a $3{\times}3$ matrix.
Now let us introduce 
the {\it standard form},
which is convenient 
for the following arguments\cite{oshikawa1}.
Choosing a unitary transformation 
${\cal U}$ appropriately,
one can obtain
$
{\cal U}{\bf H}_{0}({\bf p}){\cal U}^{-1}
=
1{\bf v}_{0}{\bf p}
+
({\sigma}_{x},
{\sigma}_{y},
{\sigma}_{z})
{\bf D}{\bf T}{\bf p}
$
where
${\bf D}$ is
${\rm diag}(1,1,{\rm sgn}({\rm det}{\bf v}))$ and
${\bf T}$ is an upper triangle matrix 
with positive diagonal elements.
Let us perform
${\bf T}{\bf p}\rightarrow {\bf p}$
(the parity-conserving  Affine transformation
on $(k_{x},k_{y})$
and the rescaling 
on $g$)
and the redefinition
${\bf v}_{0}{\bf T}^{-1}{\rightarrow}{\bf v}_{0}$.
Finally
the standard form is obtained as
\begin{eqnarray}
{\bf H}_{1}({\bf p})
&=&
1{\bf v}_{0}{\bf p}+
{\sigma}_{x}p_{x}+
{\sigma}_{y}p_{y}+
{\sigma}_{z}p_{z}{\rm sgn}({\rm det}({\bf v})).
\label{osieq3}
\end{eqnarray}
This is equivalent to 
the Hamiltonian ${\bf H}(k)$
where $A(k)=p_{z}{\rm sgn}({\rm det}{\bf v})$
and $B(k)=p_{x}-ip_{y}$.
Performing the same analysis,
one can find that 
a vortex moves from one band to the other
at the gap-closing,
and
the conclusion is that
the change of the Chern number 
is practically determined by ${\rm sgn}({\rm det}{\bf v})$
and the change is $+1$ or $-1$ \cite{oshikawa1}.
Next let us consider
a dual gap-closing point
$(-k_{x}^{0},-k_{y}^{0})$
which exists due to the symmetry.
Here the relation 
derived from (\ref{su2-rl})
plays a crucial role
and it is given by
${\bf H}(-k)=
-(\sigma_{y}{\bf H}(k)\sigma_{y})^{*}$.
Therefore
${\bf H}(-k_{x}^{0}-p_{x},-k_{y}^{0}-p_{y},g)
=-(\sigma_{y}{\bf H}(k_{x}^{0}+p_{x},k_{y}^{0}+p_{y},g)\sigma_{y})^{*}
{\sim}
-1{\bf v}_{0}{\bf p}+
({\sigma}_{x},{\sigma}_{y},{\sigma}_{z})
\bf{v}
\bf{p}.$
To summarize,
the linearized Hamiltonian
near the dual gap-closing point $(-k_{x}^{0},-k_{y}^{0})$
is given by
\begin{eqnarray}
{\bf H}(-k_{x}^{0}+p_{x},-k_{y}^{0}+p_{y},g)
{\sim}
-1{\bf v}_{0}{\bf p}
+
({\sigma}_{x},{\sigma}_{y},{\sigma}_{z})
\bf{w}
\bf{p}
\label{dual1}
\end{eqnarray}
where
${\bf w}=
{\bf v}
{\rm diag}(-1,-1,1)$.
It gives
${\rm sgn}({\rm det}{\bf v})=
{\rm sgn}({\rm det}{\bf w})$.
Therefore
the change of the Chern number
due to gap-closings
always occurs in pair with the same sign and
the total change is
${\Delta}C=\pm 2$ generally.
This is the selection rule.
On the other hand,
in the absence of the relation (\ref{su2-rl})
(or SU(2) symmetry), the above argument does not hold 
and it leads to the rule ${\Delta}C=\pm 1$.
Now we note the results
in ref.\cite{kagalovsky1} 
where the network model 
with the same symmetry as our model
was investigated.
In spite of the fact that
their model is different 
from the lattice fermion model
considered here,
our selection rule still applies:
this suggests the universality.
Assuming that
the system can belong to a phase
with a vanishing Chern number 
by tuning parameters in the Hamiltonian,
the selection rule implies that 
the Chern number is always even,
which supports the result based on the edge states.

Finally,
we comment on
the disorder-driven transition
based on 
the random $d_{x^{2}-y^{2}}+id_{xy}$ model.
It is defined by 
$t_{ij}=t_{ij}^{0}+{\delta}t_{ij}$
and ${\Delta}_{ij}={\Delta}_{ij}^{0}+{\delta}{\Delta}_{ij}$
where
${\delta}t_{ij}$ and ${\delta}{\Delta}_{ij}$
denote the randomness.
Here the hermiticity and the SU(2) symmetry
are imposed on $t_{ij}$ and ${\Delta}_{ij}$
respectively.
It has an intimate connection with 
the random Dirac fermion problem
\cite{rd1,rd1.5,rd2,rd3,rd4,fukui1}.
It is to be noted that
the SU(2) symmetry is kept
even in the presence of randomness
and the model is interesting as
the Anderson localization problem\cite{az1}. 
As discussed above,
the Chern number is ${\pm}2$
in the absence of randomness.
On the other hand,
in the presence of sufficiently strong randomness,
it is expected that 
all the vortices disappear through 
the pair annihilation of
vortices
with opposite charges
and the Chern number vanishes
\cite{hk3,hk4}.
By the numerical diagonalization,
we treated
the disorder-driven transition
for the 
${\delta}t_{ij}={\delta}_{ij}f_{i}$,
${\delta}{\Delta}_{ij}={\delta}_{ij}g_{i}$
where 
the $f$'s,$g$'s are uniform random numbers 
chosen 
from 
$[-W/2,W/2]$.
The model
was also studied extensively in ref.\cite{senthil2}.
In Fig.2,
the density of states is shown
in the case $\Delta_{xy}^{0}{\neq}0$.
It can be seen that
the two energy bands 
become closer and finally touch,
as the randomness strength is increased.
The transition
$C={\pm}2{\rightarrow}0$ 
with the gap-closing
is a natural consequence
from the selection rule.
The exploration of the global phase diagram
and the field-theoretical description
are left as future problems.

The authors thank
T.~Fukui
for valuable discussions
and P.-A.~Bares
for careful reading of the manuscript.
This work was supported in part by Grant-in-Aid
from the Ministry of Education, Science and Culture
of Japan and also Kawakami Memorial Foundation.
The computation  has been partly done
using the facilities of the Supercomputer Center,
ISSP, University of Tokyo.

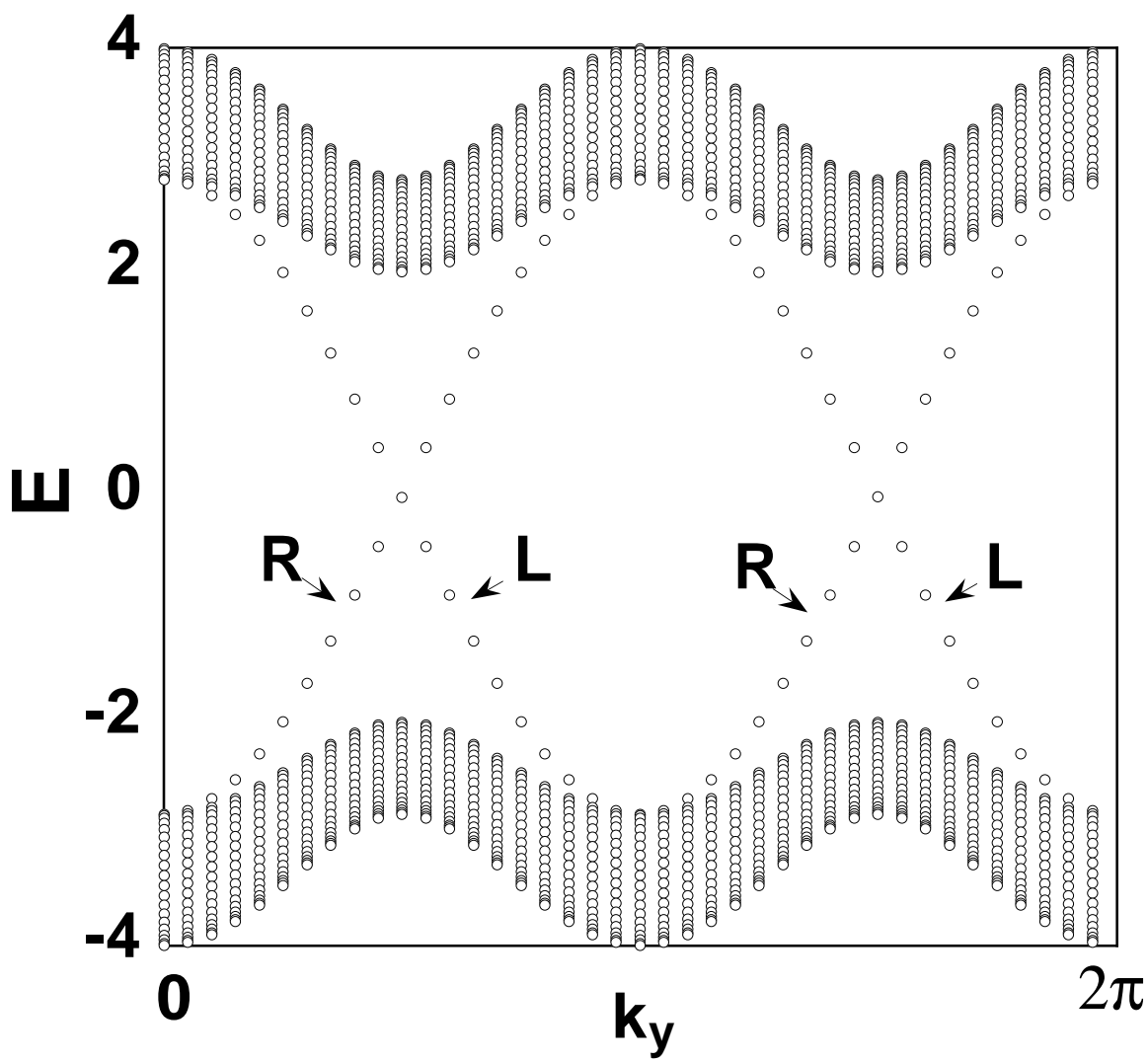
\begin{figure}
\caption{Energy spectrum 
in 
the $d_{x^{2}-y^{2}}+id_{xy}$ model
on a cylinder
($t=1$,
${\Delta}_{x^{2}-y^{2}}=1$
and ${\Delta}_{xy}=0.5$).
The system size is $40{\times}40$.
The spectrum in the energy gap
corresponds to the edge states.
The boundary where the edge states
are localized is shown by L(left) or R(right).}
\end{figure}

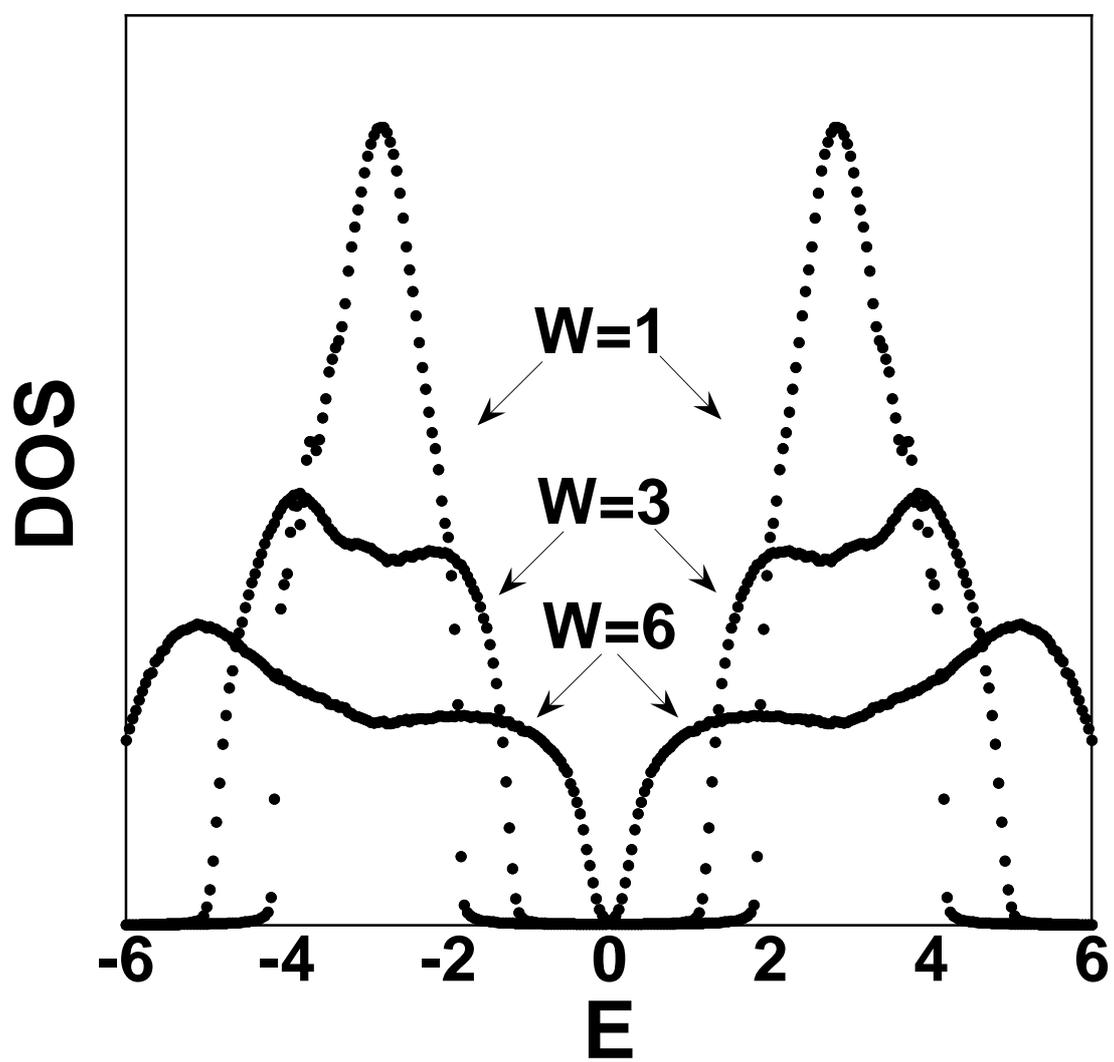
\begin{figure}
\caption{
Density of states 
in 
the random $d_{x^{2}-y^{2}}+id_{xy}$ model.
In the absence of randomness,
the model reduces to
the pure $d_{x^{2}-y^{2}}+id_{xy}$ model
with
$t=1$,
${\Delta}_{x^{2}-y^{2}}=1$,
and ${\Delta}_{xy}=0.5$.
The system size is $30{\times}30$
and the ensemble average is performed
over $800$ different realizations 
of randomness.}
\end{figure}

\end{document}